\documentclass[12pt]{article}

\usepackage{graphicx,color}
\usepackage{amsmath}
\usepackage{amssymb}
\usepackage{amsfonts}
\usepackage{latexsym}

\renewcommand{\vec}[1]{{\bf #1}}

\renewcommand{\Im}{\,\mbox{Im}\,}

\def\beq{\begin{eqnarray}}
\def\eeq{\end{eqnarray}}

\def\ln{\,\mbox{ln}\,}

\def\Res{\,\mbox{Res}\,}

\renewcommand{\Im}{\,\mbox{Im}\,}

\def\al{\alpha}
\def\be{\beta}

\def\ka{\kappa}
\def\la{\lambda}
\def\na{\nabla}
\def\pa{\partial}

\def\si{\sigma}
\def\om{\omega}

\def\th{\theta}

\def\La{\Lambda}

\def\Th{\Theta}

\renewcommand{\texttt}{{}}

\def\bs{\begin{subequations}}
\def\es{\end{subequations}}

\def\la{\lambda}

\def\om{\omega}

\newcommand{\tia}[1]{}

\begin{document}

%%%%%%%%%%%%%%%%%%%%%%%%%%%%%%%%%%%%%%
%%%%%%%%%%%%%%%%%%%%%%%%%%%%%%%%%%%%%%
%%%%%%%%%%%%%%%%%%%%%%%%%%%%%%%%%%%%%%
\begin{center}

{\large\bf  Superrenormalizable quantum gravity with complex ghosts}
%% Super-renormalizable gravity without stable ghosts}
%%%%%%%%%%%%%%%%%%%%%%%%%%%%%%%%%%%%%%
\vskip 6mm

\textbf{Leonardo Modesto$^a$}
\footnote{E-mail addtess: \ lmodesto@fudan.edu.cn}
\quad
and
\quad
\textbf{Ilya L. Shapiro$^{b,c}$}
\footnote{E-mail addtess: \ shapiro@fisica.ufjf.br}
%%%%%%%%%%%%%%%%%%%%%%%%%%%%%%%%%%%%%%
\vskip 4mm

\textsl{a) \
\small Department of Physics \& Center for Field Theory and Particle Physics,}
\\
{\small Fudan University, 200433, Shanghai, China}
\vskip 2mm

\textsl{b)}  \
{\small Departamento de Fisica - ICE,
Universidade Federal de Juiz de Fora, 33036-900
\\
Juiz de Fora, Minas Gerais, Brazil}

\vskip 2mm

\textsl{c)}  \
{\small
Tomsk State Pedagogical University and Tomsk State University
\\
634041, Tomsk, Russia}
\end{center}
\vskip 8mm

\date{\small\today}

%%%%%%%%%%%%%%%%%%%%%%%%%%%%%%%%%%%
\begin{abstract}

\noindent
We suggest and briefly review a new sort of superrenormalizable
models of higher derivative quantum gravity. The higher derivative
terms in the action can be introduced in such a way that all the
unphysical massive states have complex poles. According to the
literature on Lee-Wick quantization, in this case the theory can be
formulated as unitary, since all massive ghosts-like degrees of
freedom are unstable.
\vskip 6mm

\vskip 2mm


\end{abstract}

%%%%%%%%%%%%%%%%%%%%%%%%%%%%%%%%%%%
%%%%%%%%%%%%%%%%%%%%%%%%%%%%%%%%%%%
%%%%%%%%%%%%%%%%%%%%%%%%%%%%%%%%%%%
\section{Introduction}

One of the main theoretical problems concerning quantum gravity is a
well-known conflict between renormalizability and unitarity. Quantum
gravity theory based on general relativity is not renormalizable by
power counting, since the loop expansion parameter has inverse-mass
dimension. Renormalizability can be achieved by introducing fourth
derivative terms into the action, because in such a theory the main
coupling constant (parameter of the loop expansion in the UV) is
dimensionless \cite{Stelle-77} (see also \cite{book} for an
introduction).  However, in this case the particle
spectrum of the theory includes unphysical massive ghosts which can
not be removed without violating unitarity of the S-matrix. The
possibility to solve the problem by using the imaginary poles in the
dressed propagator of gravitons was discussed in a number of remarkable
papers \cite{Tomboulis-77,salstr,Antomb}, but the final conclusion was
that a definite solution requires full information about these
complex poles at the non-perturbative level \cite{Johnston}, that
is far away from the state of art in quantum gravity (regardless of
an interesting attempt in this direction \cite{CodPer-06}).

Introducing into the starting action some extra terms with more
than four derivatives of the metric and with real massive poles
provides a superrenormalizable theory, since the loop expansion
parameter in this case has positive mass dimension. However,
such modification does not change situation with ghosts, since
they remain in the spectrum of the theory \cite{highderi}. An
interesting possibility which should be mentioned is to introduce a
specially tuned terms which are non-polynomial in
derivatives\footnote{This was originally done for the gravitational
theories in by Tseytlin \cite{Tseytlin-95} (see also \cite{Siegel-03})
in order to provide the singularity-free modified Newtonian limit.
Recently the non-singular potential was ``rediscovered'' in
\cite{BGKM} without the use of auxiliary fields.} that can provide a
ghost-free structure of the theory at the tree level \cite{Tomboulis-97}
(see also \cite{modesto})\footnote{
More recently this theory has been generalized to any dimension
\cite{modesto2}
and explicitly showed to be finite at any order in the loop expansion
in both odd and even dimensions when some extra local operators
(only two of them are necessary in $D=4$) are included
\cite{modestoLeslaw}.}.
However, one can prove that taking loop corrections into account
the dressed propagator in such a theory gains infinitely many
ghost-like poles \cite{CountGhosts}, all of them with complex
squares of ``masses''. This situation shows that the ``ghost-free''
model of  \cite{Tseytlin-95} and \cite{Tomboulis-97}
is a non-local generalization of the historically first model
of superrenormalizable theory of quantum gravity \cite{highderi}.
The ghosts-like states are indeed present in both local polynomial
and non-local and non-polynomial versions of the theory. The
difference is that in the non-polynomial case ghosts show up
only after quantum corrections are taken into account, and that
the number of such ghosts is infinite.

In the mentioned conflict between renormalizability and
unitarity there is an unexplored possibility which we start to
consider here. Namely, in the present work we discuss a new sort
of superrenormalizable quantum gravity theory, when all massive
states correspond to the complex poles. According to existing
literature (see, e.g., \cite{LW,Veltman,Cutk,Yamamoto1}),
in this case the theory can
be formulated as unitary, since all ghosts are unstable. Due to
the existence of these works, we mainly need to review them and
discuss possible applications to higher derivative gravity models.
The organization of the manuscript is as follows.
%%%
 In Sect. \ref{Sect2} the superrenormalizable models of
 quantum gravity are briefly reviewed. After this we
consider the simplest such model with six derivatives
and obtain the conditions for the complex massive poles.
In Sect. \ref{Sect3} there is a discussion of existing works
on the Quantum Field Theories (QFT) with complex poles
and the application of complex
poles to gravity with higher derivatives. We also present an
example of how it works on a toy model with higher derivative
insertion (polynomial in the case, but in principle it can be
generalized to the general version).
In Sect. \ref{Sect4}  the stability of Lee-Wick unitarity
under the radiative corrections is discussed, for both
superrenormalizable and finite versions of higher-derivative
quantum gravity.
Finally, in Sect. \ref{Sect5} we draw our conclusions and
present some discussions.

%%%%%%%%%%%%%%%%%%%%%%%%%%%%%%
%%%%%%%%%%%%%%%%%%%%%%%%%%%%%%
\section{Superrenormalizable gravity with complex poles}
\label{Sect2}

The action of the general superrenormalizable polynomial model
\cite{highderi} can be written as
\beq
S
&=& -\frac{1}{16\pi G}
\int d^4x\sqrt{-g}\,\big(R+2\La \big)
\,+\,\int d^4x\sqrt{-g}\,\Big\{
c_1R_{\mu\nu\al\be}^2 + c_2R_{\mu\nu}^2 + c_3R^2
\nonumber
\\
&+&
d_1 R_{\mu\nu\al\be} \Box R^{\mu\nu\al\be}
+ d_2 R_{\mu\nu}\Box R^{\mu\nu}
+ d_3 R\Box R
\, + \,d_4R^3 \, + \,d_5 R R^{\mu\nu}R_{\mu\nu} \,+\,\dots
\nonumber
\\
&+&
\,\dots
\nonumber
\\
&+&
f_1 R_{\mu\nu\al\be} \Box^k R^{\mu\nu\al\be}
+ f_2 R_{\mu\nu}\Box^k R^{\mu\nu}
+ f_3 R\Box^k R\,+\, \dots\,+\, f_{4,5,..}R_{...}^{k+2}
\Big\}\,.
\label{superre}
\eeq
Here the first integral is the Einstein-Hilbert action with
cosmological constant and the second includes higher
derivative terms.
We assume that  $k=1,2,\dots\,$. The terms indicated by
dots in (\ref{superre}) and the terms $f_{4,5,..}R_{...}^{k+2}$
denote the set of all covariant local terms with the derivatives
up to the order $2k+4$.  All surface terms are omitted for brevity.
$\,c_{1,2,3},\,d_{1,2,3},\,...f_{1,2,\,...}\,$ are arbitrary
coefficients.

The discussion of unitarity and renormalization is much simpler
for the flat background, hence in what follows we assume that
$\La=0$. As it was explained already in \cite{Stelle-77}, the
results are not affected by this assumption.

The evaluation of the superficial degree of divergence $D$ of
the Feyman diagrams in the theory (\ref{superre}) leads to the
 following result\footnote{This expression corresponds to the
 four spacetime dimensions. Generalization to an arbitrary
dimension is possible, but it will be considered elsewhere
\cite{LMso}.} \cite{highderi}:
\beq
D\,+\,d &=& 4\,+\,k(1-p)\,,
\label{sup-pow}
\eeq
where $d$ is the number of metric derivatives in the counterterms
at the $p$-loop level. For the logarithmically divergent diagrams
with $D=0$ the relation (\ref{sup-pow}) indicates that the models
with $k=1$ have divergences only up to the three-loop order, for
$k=2$ divergences show up only up to the two-loop order. Finally,
the  models with $k \geq 3$ may have only one-loop divergences
with restricted number of derivatives of the metric, $d=4$, $d=2$
and $d=0$. Moreover,
in all cases only the parameters $\La,\,G,\,c_{1,2,3}$ gain
divergent contributions, and therefore the coefficients of the
higher derivative terms do not require infinite renormalization.
This means also that the terms with derivatives higher than
four are not running. At the same time, the coefficients of the
highest derivative terms define the running of the cosmological
and Newton constants and of the coefficients $c_{1,2,3}$.
For $k \geq 3$ the corresponding one-loop beta-functions
are exact.

In order to complete the story, let us note that one can choose the
terms with highest derivatives and terms ${\cal O}(R^3_{\dots})$
in such a way that the divergences cancel and the theory becomes
finite \cite{modesto} (see also \cite{CountGhosts} for an alternative
consideration).  Furthermore, if there are divergences, they do not
depend on the choice of the gauge-fixing parameters, hence the
$\be$-functions in this theory are unambiguous.

Indeed, the renormalizability or superrenormalizability of the
theories (\ref{superre}) has a price, and this price is not small.
The physical spectrum of the theory includes not only a usual
massless graviton, but also a set of massive tensor and scalar
modes, and part of these extra degrees of freedom are ghosts.
In the first paper \cite{highderi} it was shown that the models with
real mass spectrum always have both ghosts and healthy massive
fields, with alternating signs of the masses and residues. In the
present work we shall elaborate on the case of complex masses of
the ghost modes.

In what follows we shall discuss general $k$, but will mainly
concern describing the simplest situation with complex poles
for the model with $k=1$. The situation for $k > 1$ is qualitatively
similar and hence the simplest $k=1$ case gives sufficiently clear
general understanding.
The structure of poles in the propagator on a flat background
is defined by the terms which are at most quadratic in curvature
tensor. One can make further simplification if remember the
relation
\beq
R_{\mu\nu\al\be} \Box^l R^{\mu\nu\al\be}
- 4 R_{\mu\nu}\Box^l R^{\mu\nu}
+ R\Box^l R
&=&
\na_\mu \chi^\mu
\,+\,{\cal O}\big(R_{...}^3\big)\,,
\quad \mbox{ for any}\,\, l\,.
\label{GBmod}
\eeq
As a result the analysis of the propagator can be done for
$c_1=d_1 = ... =f_1=0$. It proves useful to introduce another
basis and notations for the relevant terms in the action
\beq
S_{red}
&=& - \frac{2}{\ka^2}
\int d^4x\sqrt{-g}\,R
\nonumber
\\
&-&
\al \int d^4x\sqrt{-g}\,\Big\{
\frac12\,C_{\mu\nu\al\be}\,
\Pi_2\big(\Box\big)\,C^{\mu\nu\al\be}
\,+\,
\om\,R\,
\Pi_0\big(\Box\big)\,R
\Big\}\,,
\label{supred}
\eeq
where $C_{\mu\nu\al\be}$ is Weyl tensor,
$\ka=(32\pi G)^{-1/2}=1/M_P$ is the inverse of reduced
Planck mass, $\,\al\,$
and $\,\om\,$ are arbitrary numerical parameters and
$\Pi_{2,0}(x)=1+\,...$
are some polynomials of order $k$. With this notations
we can use directly the results of \cite{Stelle-78} to arrive at
the part of the action (\ref{supred}) which is quadratic in
the perturbations\footnote{Let us
note that there is a gauge fixing dependence in the scalar sector
of this expression, which was discussed in \cite{book}. However,
the expression (\ref{supred2}) which we take from \cite{Stelle-78}
corresponds to the gauge-independent interaction between two
sources and hence can be considered as well-defined at the
tree-level.},
$\,\ka h_{\mu\nu}=g_{\mu\nu}-\eta_{\mu\nu}$,
\beq
S^{(2)}_{red}
&=&
-\,\int d^4x
\Big\{
\frac12\,h^{\mu\nu}\Big[\frac{\al\ka^2}{2}
\Pi_2\big(\pa^2\big)\pa^2-1\Big]\,\pa^2
\,P^{(2)}_{\mu\nu,\,\rho\si}\,h^{\rho\si}
\nonumber
\\
&+&
h^{\mu\nu}\Big[\al\om\ka^2\Pi_0\big(\pa^2\big)\pa^2-1\Big]
\,\pa^2\,P^{(0-s)}_{\mu\nu,\,\rho\si}\,h^{\rho\si}
\Big\}\,.
\label{supred2}
\eeq
An obvious difference between (\ref{supred}) and (\ref{supred2})
is that the last is based on the flat-space metric and on the partial
derivatives (e.g., $\pa^2=\eta^{\mu\nu}\pa_\mu\pa_\nu$), while
 the first is a covariant expression. The projectors to tensor
$P^{(2)}_{\mu\nu,\,\rho\si}$ and scalar $P^{(0s)}_{\mu\nu,\,\rho\si}$
states on the flat background are defined in a standard
way (see, e.g., \cite{Stelle-77} or \cite{book}),
\beq
&&
P^{(0-s)}_{\mu\nu,\,\rho\si}
=\frac13\,\th_{\mu\nu}\,\th_{\rho\si}
\,,\quad
P^{(2)}_{\mu\nu,\,\rho\si}
= \frac12\,\big(\th_{\mu\rho}\,\th_{\nu\si}
+ \th_{\nu\rho}\,\th_{\mu\si}\big)
- P^{(0-s)}_{\mu\nu,\,\rho\si}\,,
\nonumber
\\
&& \mbox{where}
\quad
\th_{\mu\nu} = \eta_{\mu\nu}-\frac{\pa_\mu\pa_\nu}{\pa^2}\,.
\label{project}
\eeq

After Wick rotation to Euclidean space, the equations
for the poles have the form
\beq
&&
\al\Pi_2 (p^2)p^2\,=\,2M_P^2
\,,\quad
\al\om\,\Pi_0(p^2)p^2\,=\,M_P^2\,.
\label{poles-geral}
\eeq
In the simplest case of the fourth-derivative theory
\cite{Stelle-77,Stelle-78}, $\Pi_2=\Pi_0=1$, hence the solutions
for the poles, in the tensor and scalar sectors, are
\beq
&&
p^2 = m_2^2 = \frac{2M_P^2}{\al}
\quad
\mbox{and}
\quad
p^2 = m_0^2 = \frac{M_P^2}{\al\om}\,.
\label{poles-four}
\eeq
The positive signs of the masses corresponds to the negative
sign of the higher-derivative terms in (\ref{supred}).

Let us consider the next order and choose
\beq
&&
\Pi_2 (p^2)\,=\,1 + \frac{p^2}{2A_2}
\,,\qquad
\Pi_0(p^2)\,=\,1 + \frac{p^2}{2A_0}\,,
\label{poly-six}
\eeq
where $A_0$ and $A_2$ are some constants with the dimension of the
square of mass. Since the two equations (\ref{poles-geral}) are
similar, let us present the solution only for the tensor part,
\beq
&&
p^2 = m_2^2 = - A_2 \pm \sqrt{A_2^2 + \frac{4A_2M_P^2}{\al}}\,.
\label{poles-six-geral}
\eeq
This expression shows that, in principle, one can have the
following types of solutions:
\vskip 2mm

\noindent
{\large \textbullet} \ \
Two real positive solutions $0<m_{2a}^2<m_{2b}^2$ for the poles.
This is the case discussed in \cite{highderi} and we will not
consider it again here.
\vskip 2mm

\noindent
{\large \textbullet} \ \
Two pairs of complex conjugate solutions for the mass, one with positive
and one with negative imaginary parts. In the rest of the paper we
concentrate on this case, assuming that the poles of the propagator
include one massless state (graviton) and that all other poles are
massive  and complex.

The quantization of QFT with
complex poles was pioneered by Lee and Wick in \cite{LW}.
The subject attracted a great deal of attention see, e.g.,
\cite{Veltman,Cutk,Yamamoto1}. The net result is that such theories
may be formulated as unitary, since the optical theorem (see, e.g.,
\cite{BogShirk,Peskin,Schwartz}) is satisfied\footnote{An interesting
attempt to implement this scheme in the four-derivative
quantum gravity can be found in \cite{Narain}, but there is
no concrete mechanism which provides an absence of physical
poles in this case.}.
The complex conjugate poles
do not appear on shell since this would mean that one meets
physical observables with imaginary component. The physically
relevant part of the propagator is composed only by the states
corresponding to real poles. In our case this means that the
physically relevant part of the propagator is the same as in
Einstein gravity.

The application of these ideas to the fourth-derivative Quantum
Gravity also has a long history, starting from the works of
Stelle \cite{Stelle-77}, Tomboulis \cite{Tomboulis-77}
and Salam and Strathdee \cite{salstr}, where the condition of
unitarity at the quantum level has been formulated in form of
the Froissart Bound which must be satisfied by a dressed
propagator with quantum corrections taken into account. The
one-loop corrections typically split the real massive pole
into a couple of complex conjugate poles
\cite{Tomboulis-77,HasMottola-81,Antomb,Hawking}.
The explicit (rather complicated, technically) calculations of the
one-loop  corrections in four-derivative gravity were done in
\cite{julton,frts82,avrbar} and carefully checked in \cite{Gauss},
including the hypothesis of the relevant role of the Gauss-Bonnet
term \cite{capkim}. We know that the one-loop $\beta$-functions
\cite{frts82,avrbar} have ``correct'' signs, exactly as the
contributions of matter fields, which can be also used in the
framework of the large-$N$ approximation \cite{Tomboulis-77}.
Unfortunately, it was shown in Ref. \cite{Johnston} that the
one-loop calculations are not completely conclusive, the same
also concerns the large-$N$ approximation, which does not provide
reliable non-perturbative information about quantum gravity.
Hence the question of whether the dressed propagator of
metric perturbations has a form which satisfies the Froissart
Bound remains open, until we will be able to get a full
non-perturbative form of quantum corrections.

From our
point of view the great advantage of the superrenormalizable
models (\ref{superre}) is that in this case one can provide
the desirable complex structure of the propagator already at the
tree-level. In this case the conditions of Ref. \cite{salstr}
can be easily satisfied. Moreover, in the superrenormalizable
theory the form of loop corrections can be easily set under
control, hence the desirable form holds also for the dressed
propagator with full quantum corrections.  In the next section
we present a brief description of the unitarity in the theory
with complex poles at the tree level and after that discuss the
loop corrections.

%%%%%%%%%%%%%%%%%%%%%%%%%%%%%%
%%%%%%%%%%%%%%%%%%%%%%%%%%%%%%
\section{Unitarity in the theory with complex poles}
\label{Sect3}

In QFT unitarity of the $S$-matrix means
\beq
S^\dagger S = 1 \,.
\label{uni}
\eeq
In terms of the $T$-matrix defined by
\beq
S = 1 + i T \,,
\eeq
the unitarity condition (\ref{uni}) turns out to be
\beq
- i ( T - T^\dagger) = T^\dagger T\,.
\eeq
One has to consider the matrix element of the above equation
between the initial state $|i \rangle$ and the final state
$\langle f|$,
\beq
- i \left(   \langle f | T | i \rangle  - \langle f | T^\dagger | i \rangle  \right)
&=&
\langle f|T^\dagger \Big( \sum_k |k \rangle \langle k|\Big) T|i \rangle\,,
\nonumber
\eeq
By defining the scattering amplitude as
\beq
\langle f | T | i \rangle
&=&
(2 \pi)^D \delta^{D}(p_i - p_f) \, T_{f i} .
\eeq
we arrive at
\beq
- i \left( T_{f i} - T^*_{i f} \right)
&=&
\sum_k  T^*_{k f} T_{k i}\,.
\label{matrixT}
\eeq
Assuming that for the forward scattering amplitude $i = f$,
%% and
%% that the theory is invariant under the inversion
%% $x^\mu \rightarrow - x^\mu$,
%% $T_{i f} = T_{fi}$,
previous equation simplifies to
\beq
2 \,  \Im T_{ii}
&=&
\sum_k T^*_{i k} \, T_{ik}  > 0 \,.
\label{Tunitarity}
\eeq

We now present a systematic consideration of the tree-level unitarity,
partially following the work by Accioly et al, Ref. \cite{Accioly-2002}.
A general theory is certainly well-defined if  ``tachyons" and ``ghosts"
are absent, in which case the propagator has only first poles at
$k^2 - M^2 =0$ with real masses (no tachyons) and with positive
residues (no ghosts). In order to test the tree-level unitarity
of a superrenormalizable higher derivative gravity one can
introduce an external conserved stress-energy tensor,
$\Theta^{\mu \nu}$, and examine the amplitude at the pole.
When we introduce such a general source, the linearized action
including the gauge-fixing term reads
\beq
\mathcal{L}_{h\Theta}
&=&
\frac{1}{2} h^{\mu\nu} \mathcal{O}_{\mu\nu ,\rho \si} h^{\rho\si}
- g\,h_{\mu \nu} \Theta^{\mu \nu}\,.
\label{LGM}
\eeq

The transition amplitude in momentum space is defined by the
expression
\beq
i T
&=&
(- i)^2 g^2 \, \Th^{\mu\nu}\, i \Delta_{F \mu \nu , \rho\si}
\, \Th^{\rho \si} \, ,
\nonumber
\\
\mbox{where}
&&
\langle 0 | T \big\{ h_{\mu\nu}(x') h_{\rho\si}(x) \big\} |0 \rangle
\,=\,
i \Delta_{F \mu\nu, \rho\si} (k)
\,\equiv\, i \mathcal{O}^{-1}_{\mu \nu , \rho \sigma} (k) \,.
\label{ampli1}
\eeq
It proves useful to expand the sources using
independent vectors in the momentum space,
\beq
k^{\mu} = (k^0, \vec{k})
\,, \quad
\tilde{k}^{\mu} = (k^0, - \vec{k})
\,, \quad
\epsilon^\mu_i = (0, \boldsymbol{\epsilon}_i)
\,, \quad
i =1,2 \,,
\eeq
where $\boldsymbol{\epsilon}_i$ are unit vectors orthogonal to each
other and to $\vec{k}$. The symmetric stress-energy tensor reads
\beq
\Theta^{\mu\nu}
&=& a k^{\mu} k^{\nu} + b \tilde{k}^{\mu} \tilde{k}^{\nu}
+ c_{i j} \epsilon_i^{(\mu} \epsilon_j^{\nu)}
+ d \, k^{(\mu} \tilde{k}^{\nu)} + e_i k^{(\mu} \epsilon_i^{\nu)}
+ f_i \tilde{k}^{(\mu} \epsilon_i^{\nu)}\,,
\label{20}
\eeq
where  $a,b,c_{ij},d, e_i, f_i$ are some coefficients, which can be
partially constrained by the conservation law conditions
$k_{\mu} \Theta^{\mu \nu} = 0$.

In the presence of the usual graviton pole and a finite sequence of
complex conjugate poles the Feynman propagator reads
\beq
i \Delta_F(k) = i \Big[
\frac{1}{k^2 + i \epsilon} + \sum_n \Big( \frac{c_n}{k^2 - \eta_n^2}
+ \frac{c^*_n}{k^2 - {\eta_n^*}^2} \Big) \Big]
\Big( P^{(2)} - \frac12\,P^{(0)}\Big)\,,
\label{propCC}
\eeq
where the projectors $P^{(2)}$ and $P^{(0)}$ can be consulted
in Eq. (\ref{project}). In the
last formula the spacetime indexes in $\Delta_F$ and in the
projectors are omitted\footnote{For the sake of simplicity in
(\ref{propCC}) we considered a special higher derivative theory
with polynomials of $\Box$  present only between the Einstein
tensor $G_{\mu \nu}$ and the Ricci tensor $R_{\mu \nu}$.}
Replacing the last expression into
(\ref{ampli1}) we arrive at the result,
\beq
i T
&=&
= (2\pi)^D \delta(P_i - P_f) \, i \, T_{i f}
\nonumber
\\
&=& - g^2 \,
\Theta^{\mu \nu} i \Delta_{F \mu \nu , \rho \sigma} \Theta^{\mu \nu}
\,=\,(2\pi)^D \delta(P_i - P_f) \, i \,T_{if}\,,
\label{ampli1-2}
\eeq
where
\beq
 T_{i f} =
(- i)^2 \, \Theta^{\mu\nu}
\Big[
\frac{1}{k^2 + i \epsilon} + \sum_n \Big( \frac{c_n}{k^2 - \eta_n^2}
+ \frac{c^*_n}{k^2 - {\eta_n^*}^2} \Big) \Big]
\Big( P^{(2)} - \frac12\,P^{(0)}\Big)_{\mu\nu,\rho\si} \Theta^{\rho\si} \,.
\label{ampli1-3}
\eeq

Using projectors (\ref{project}) and the conservation low
$k_{\mu} \Theta^{\mu \nu} = 0$ in (\ref{ampli1-3}), the
imaginary part of $T_{if}$ reads
\beq
\Im \,T_{i f}
&=&
\Im \,
(- ig)^2  \Big(\Theta_{\mu \nu} \Th^{\mu\nu}
- \frac12\,{\Th^\mu_\mu}^2 \Big)\,
\Big[
\frac{k^2 - i \epsilon}{k^4 + \epsilon^2}
+ \sum_n \Big( \frac{c_n}{k^2 - \eta_n^2} + \frac{c^*_n}{k^2 - {\eta_n^2}^*}
\Big) \Big]
\nonumber
\\
&=&
\frac{ g^2 \,\epsilon}{k^4 + \epsilon^2}
\,\Big( \Th_{\mu\nu} \Th^{\mu\nu} - \frac12\, {\Th^\mu_\mu}^2
\Big)
\,\longrightarrow\,
\pi  g^2\, \Big( \Th_{\mu\nu} \Th^{\mu\nu}
- \frac12\, {\Th^\mu_\mu}^2 \Big)
\, \delta(k^2)\,,
\label{ampli2}
\eeq
where the usual cut rule has been assumed at the limit $\epsilon \to 0$.
From (\ref{Tunitarity}) and (\ref{ampli2}) the tree-level unitarity
requirement simplifies to
\beq
\Im \Big\{
 \Theta(k)^{\mu\nu} \mathcal{O}^{-1}_{\mu\nu, \rho \sigma}
 \Theta(k)^{\rho \sigma} \Big\}
 \,=\,
\pi \, \Res
  \Big\{
 \Theta(k)^{\mu\nu} \mathcal{O}^{-1}_{\mu\nu, \rho \sigma}
 \Theta(k)^{\rho\si} \Big\}_{k^2 = 0}> 0
 \eeq
and (\ref{ampli2}) can be recast into
\beq
\Res \left( \mathcal{A} \right) \big|_{k^2 =0}
&=&
g^2 \Big( c_{ij}^2 - \frac12\,c_{ii}^2 \Big)\,,
\label{residuo}
\eeq
with the coefficients $c_{ij}$ defined in (\ref{20}).

In the Lee-Wick theory the propagator shows extra complex poles
and at the moment it is not obvious how to derive, if any, the usual
Largest Time Equation. However, we can still analyze Eq.
(\ref{Tunitarity}) for the case of individual graphs by cutting the
diagrams \cite{Cutk} (see also \cite{Schwartz} for the introduction).
Energy-momentum conservation must be satisfied by both sides of
(\ref{Tunitarity}). Therefore, if we cut through normal particle
propagators (in our case this means the massless graviton) we have
to replace the propagator with $\delta(k^2)$. If we cut trough
the Lee-Wick propagators, these just correspond to take the imaginary
part of the sum in (\ref{propCC}), and the imaginary part of the sum
of complex conjugate poles vanish. In particular, in $T^\dagger T$
we only have to sum over intermediate normal particle states.
Therefore, the theory is unitary in the subspace of the real normal
and stable particles as a consequence of the energy-momentum
conservation and the presence of extra poles in the propagator
that always come in complex conjugate pairs.
%%%%%%%%%%%%%%%%%%

In order to illustrate the general arguments, let us consider, as a
toy  model, the case of a relatively simple Lee-Wick interacting
theory to explicitly show the perturbative unitarity of theories
with complex conjugate poles. Consider a theory of  scalar field
with cubic interaction $\la \phi^3$.   The one-loop self-energy
diagram provides the contribution to the $S$-matrix as follows:
\beq
S^{(2)} = i T = (- i)^2 \lambda^2 \int \frac{d^4 k}{(2 \pi)^4}
\,\frac{i}{(k^2 + i \epsilon) \Pi(k^2)}\,\cdot \,
 \frac{i}{[(k+p)^2 + i \epsilon] \Pi((k+p)^2)}\,,
\eeq
where $\Pi(k^2)$ has only complex conjugate zeros. For further simplicity,
 one can take an example with one pair of complex conjugate poles,
 $\Pi(k^2)= 1+ k^4/\Lambda^4$.
To verify the unitarity condition (\ref{matrixT}) we can extract the
imaginary part of the integral above without evaluating it explicitly,
\beq
\Im  T
 &=&
\Im
\Big\{ - i  \lambda^2
\,\,
\int \frac{d^4 k}{(2 \pi)^4} \frac{1}{(k^2 + i \epsilon) \Pi(k^2)}
\,\cdot\, \frac{1}{[(k+p)^2 + i \epsilon] \Pi((k+p)^2)} \Big\}
 \nonumber \\
 %\eeq
% \beq
&=&
 \Im  \Big\{
 \!\! - i  \lambda^2 \!\! \int \!\! \frac{d^4 k}{(2 \pi)^4}\,
 \frac{ k^2 (k+p)^2  - \epsilon^2 -  i \epsilon k^2
 -  i \epsilon (k+p)^2}{(k^4 +  \epsilon^2) \Pi(k^2)\,
 \cdot\, [(k+p)^4 + \epsilon^2]  \Pi((k+p)^2)}
\Big\}
\nonumber \\
 %\eeq
 %\beq
 &=&
 \Im
 \bigg\{
\!\!
- i   \lambda^2 \!\! \int \frac{d^4 k}{(2 \pi)^4}
\bigg[
\frac{k^2}{(k^4 +  \epsilon^2) \Pi(k^2)}
 \,\cdot\,\frac{(k+p)^2 }{[(k+p)^4 +  \epsilon^2] \Pi((k+p)^2}
\nonumber
 \\
&-&
 \frac{ \pi  \delta(k^2) }{ \Pi(k^2)}
\,\cdot\,\frac{\pi \delta( (k+p)^2) }{\Pi((k+p)^2)}
+  \frac{k^2 }{(k^4 +  \epsilon^2) \Pi(k^2)}
    \,\cdot\,\frac{ - i \pi \delta((k+p)^2)}{ \Pi((k+p)^2)}
\nonumber
\\
&+&
\frac{- i \pi \delta(k^2) }{ \Pi(k^2)}\,\cdot\,
 \frac{(k+p)^2}{[(k+p)^4 + \epsilon^2]\Pi((k+p)^2)}\bigg]\bigg\}
 \nonumber \\
% \eeq
 %%%%%%%%%%%%%%%%%%%%%%%%%%%%
 %%%%%%%%%%%%%%%%%%%%%%%%%%%%
 %%%%%%%%%%%%%%%%%%%%%%%%%%%%
 %\beq
  &=&
  \Im  \bigg\{ \!\!  - i  \lambda^2 \!\!
  \int \frac{d^4 k}{(2 \pi)^4}
  \bigg[ \frac{k^2 }{(k^4 +  \epsilon^2) \Pi(k^2)}\,\cdot\,
 \frac{(k+p)^2 }{[(k+p)^4 +  \epsilon^2] \Pi((k+p)^2)}
\nonumber
\\
&-&
\pi^2  \delta(k^2)
\, \delta( (k+p)^2)
+ \frac{k^2 }{(k^4 +\epsilon^2)\Pi(k^2)}(- i) \pi \delta((k+p)^2)
\nonumber
\\
&+&
 (- i) \pi \delta(k^2)
 \frac{(k+p)^2  }{[(k+p)^4 + \epsilon^2]\Pi((k+p)^2)} \bigg] \bigg\}
 \,.
\label{1loop}
\eeq
The imaginary parts of the last two integrals are zero, while the
second integral is identical to the right hand side of equation
(\ref{matrixT}). Then we still have to prove that the imaginary
part of the fist integral is zero. The expression is
%% Let us concentrate on the first integral in (\ref{1loop}).
\beq
  &&\hspace{-1cm}
 \Im \bigg\{
 - i  \lambda^2 \!\!
 \int \frac{d^4 k}{(2 \pi)^4}  \,\frac{k^2 }{(k^4 +  \epsilon^2) \Pi(k^2)}
 \,\cdot\,\frac{(k+p)^2 }{[(k+p)^4 +  \epsilon^2] \Pi((k+p)^2)} \bigg\}\,.
\label{1loopM}
\eeq
To evaluate (\ref{1loopM}) one can make the Wick rotation
$k_0 = i k_4$, such that $\epsilon$ can be discorded. Then the
imaginary part of the integral reads
\beq
\Im
\bigg\{  - i  \lambda^2 \!\! \int
\,\,i\,
\frac{d^4 k_E }{(2 \pi)^4}\, \frac{1}{k_E^2 \, \Pi(k_E^2)} \,\cdot\,
 \frac{1}{(k_E+p_E)^2  \, \Pi((k_E+p_E)^2)} \bigg\} \,=\, 0 ,
 \label{1loopE}
\eeq
and it is zero because of the extra factor of $i$.
One can find the same result using the Feynman $\epsilon$ prescription.
We end up with the following amplitude,
\beq
\Im  T
&=&
\pi^2  \lambda^2 \!\! \int   \frac{d^4 k }{(2 \pi)^4}
 \, \delta(k^2)  \,\, \delta( (k+p)^2) \,.
  \label{Ddeltas}
\eeq
which is exactly the right side of (\ref{matrixT}).

Notice that if $\Pi(k^2)$ has a pole of complex mass square (without
complex conjugate partner) then the integral (\ref{1loopE}) acquires
an imaginary part. If we have derivative self-adjoint interactions
only even powers of the momenta
$k$ will give a non-zero contribution to the loop integrals, and
the proof of this section remains basically unchanged. In other
words unitarity in a theory with complex conjugate poles works
like in a two derivative theory.

The generalization to the case of higher number of loops is
straightforward. Taking the imaginary part of any amplitude,
we get the right product of Dirac delta, namely the generalization
of (\ref{Ddeltas}) to the case of a given loop order,  plus
extra integrals whose contribution to the imaginary part is
identically zero.

%%%%%%%%%%%%%%%%%%%%%%%%%%%%%%%%
%%%%%%%%%%%%%%%%%%%%%%%%%%%%%%%%
\section{Quantum corrections and complex poles}
\label{Sect4}

At quantum level the theory can be superrenormalizable or it can be
even finite, if terms of third- and fourth-order in curvature are
properly introduced (see details in \cite{CountGhosts}). In both
cases there will be a non-local form factor in the propagator of
gravitational field at the quantum level. The difference is that
for the finite theories the beta functions are zero, there is no
need to introduce counterterms, and the propagator does not gain
logarithmic form factors, that means the non-local form factors
are weaker that logarithmic. In the finite theories of massless fields
this should mean that there will not be any quantum corrections
to the propagator. However in the superrenormalizable quantum
gravity there are mass parameters, hence this scenario is not possible.

From the general perspective it is somehow simpler to evaluate
the effect of quantum correction in the case where divergences
and logarithmic form factors occur, but when this happens only
at the one-loop level. For the sake of simplicity, let's consider
the six-derivative version with one-loop divergences only.

According to the power counting in Eq. (\ref{sup-pow}), in
this case the modification in the first of equations of
(\ref{poles-geral}) and (\ref{poly-six}) are as follows
\footnote{The consideration for the scalar part is very
similar and we will not bother the readers with repetition.}
\beq
&&
\al\Big[1 + \be_W \ln\Big(\frac{p^2}{\mu^2}\Big)
\,+\,\frac{p^2}{2A_2}\Big]p^2
\,=\,2M_P^2 \Big[ 1 + \be_\ka \ln\Big(\frac{p^2}{\mu^2}\Big)\Big]\,.
\label{poles-geral-q}
\eeq
One can see that the six-derivative term does not gain logarithmic
quantum correction, while such contributions are present for the
four-derivative and Einstein-Hilbert terms. The values of the
beta-functions $\be_\ka$ and $\be_W$ depend on the relations
between highest derivative terms and terms which are cubic and
fourth-order in curvature. One can provide desirable values
to $\be_\ka$ and $\be_W$ without much effort. For the sake of
simplicity we assume here that the beta-function for the
cosmological constant is identically zero. This is also
easy to provide, just looking at the expression which was
derived explicitly in \cite{highderi}.

The positions and properties of the poles depend on the of the
solutions of Eq. (\ref{poles-geral-q}). From the mathematical side
it is a complicated transcendental equation which can not be solved
analytically. However, this equation enables one to easily make a
qualitative analysis, which can be also confirmed by numerical
study with some given values of the parameters.
In order to understand the role of the quantum corrections in Eq.
(\ref{poles-geral-q}), let us first note that the logarithmic
form of the form factors actually holds only in the far UV, when
the energy scale is much higher that the masses of the fields.
In the present case, when the masses are defined by the Planck
scale, this limit means a far transplackian energies. On the
contrary, at the lower energies, comparable to and below the
Planck mass, the effect of masses starts to be essential.
At the sub-Planck energies we are probably going to observe a
kind of quantum decoupling, such as it was obtained for the
semiclassical corrections to gravity from massive fields
\cite{apco}. The concrete form of the form-factors is relatively
complicated, but the general structure is similar to the simple
replacement (using Euclidean signature)
\beq
\ln\Big(\frac{p^2}{\mu^2}\Big)
\,\to \, \ln\Big(\frac{p^2+m^2}{\mu^2}\Big)
\label{IR}
\eeq
 in (\ref{poles-geral-q}). At this point we can make the following
 consideration. The logarithmic functions in  (\ref{poles-geral-q})
 are slowly varying everywhere except the IR regime, where they
 must be replaced to other even more slowly varying functions
 qualitatively similar to (\ref{IR})\footnote{Situation is going to be
 very similar for the finite theories, where we also have only weaker
 than logarithmic corrections.}. This means that the non-local
 logarithmic insertions in the last equation do not increase the
number of poles in the propagator, they can only lead to a certain
shift in the positions of existing complex poles. The situation is
opposite to the one in the theory with real massive poles, since in
this case logarithmic corrections lead to she splitting of massive
real poles into a couple of complex conjugate poles
\cite{Tomboulis-77,salstr,HasMottola-81}.

After all, the Lee-Wick - type unitarity is save in the finite or in
general superrenormalizable theories of gravity which are based
on the polynomial actions (\ref{superre}).

%%%%%%%%%%%%%%%%%%%%%%%%%%%%%%%
%%%%%%%%%%%%%%%%%%%%%%%%%%%%%%%
\section{Conclusions and discussions}
\label{Sect5}

The semiclassical or quantum treatment of gravity in four-dimensional
space-time always leads to higher derivatives in the action.  In the
minimal four-derivative version this means either massive unphysical
ghost or tachyon. Including more derivatives with extra massive
{\it real} poles in the propagator does not change this fact and does
not make ghosts more massive \cite{highderi}, since ghost is always a
lightest massive spin-2 excitation, after the massless graviton.

In this paper we introduced a new type of the superrenormalizable
theories of gravity which are based on the polynomial actions
(\ref{superre}). Different from the models considered before in
\cite{highderi}, this new type of theories does not have a hierarchy
of massive ghost and massive normal particles with growing masses.
Instead, the massive poles appear in a complex conjugate pairs.
The quantization of the theories with complex poles is well-known
\cite{LW,Veltman,Cutk,Yamamoto1}, but has been never applied
to the important case of higher derivative gravity. Our analysis
shows that there is a chance to formulate the theory of gravity
with complex poles as unitary in the  Lee-Wick approach.

The introduction of six- and higher-derivative terms into the
action can be seen as an UV completion of the theory with four
derivatives, which is useful to remove physical real massive
poles from the spectrum. We do
not pretend to say the final word on this subject here. But, in our
opinion the theories with complex poles at the classical level
deserve complete study. In case of a completely consistent
formulation, the theory (\ref{superre}) with complex massive
poles may be an ideal starting point to construct a successful
theory of quantum gravity.

\section*{Acknowledgements}

Authors are grateful to Leslaw Rachwal for useful observations. 
I.Sh. is also grateful to CNPq,  FAPEMIG and ICTP for partial
support of his work.

%%%%%%%%%%%%%%%%%%%%%%%%%%%%%%
%%%%%%%%%%%%%%%%%%%%%%%%%%%%%%

\end{document}